\documentclass[seceq]{ptptex}

\usepackage{graphicx}

\title{
Local virial relation and velocity anisotropy for collisionless 
self-gravitating systems
}

\author{
Yasuhide \textsc{Sota}$^{1,2}$,
Osamu \textsc{Iguchi}$^{1}$,Masahiro \textsc{Morikawa}$^{1}$
and Akika \textsc{Nakamichi}$^{3}$
}

\inst{
$^1$ Department of Physics, Ochanomizu University,
2-1-1 Ohtuka, Bunkyo-ku, Tokyo,112-8610, Japan \\
$^{2}$ Advanced Research Institute for Science and Engineering,
Waseda University, Ohkubo, Shinjuku--ku, Tokyo, 169-8555, Japan \\
$^3$ Gunma Astronomical Observatory,
6860-86, Nakayama, Takayama, Agatsuma, Gunma, 377-0702, Japan \\
}

\abst{
The collisionless quasi-equilibrium state realized 
after the cold collapse of self-gravitating systems 
has two remarkable characters. 
One of them is the linear temperature-mass (TM) relation, 
which yields a characteristic non-Gaussian velocity distribution. 
Another is the local virial (LV) relation,
the virial relation which holds even locally in collisionless systems 
through phase mixing such as cold-collapse. 
A family of polytropes are examined from a view point of these two characters.
The LV relation imposes a strong constraint on these models: 
only polytropes with index $n \sim 5$ 
with a flat boundary condition at the center
are compatible with the numerical results, 
except for the outer region.
Using the analytic solutions based on the static and spherical Jeans equation, 
we show that this incompatibility in the outer region 
implies the important effect of  anisotropy of velocity dispersion.
Furthermore, the velocity anisotropy is essential
in explaining various numerical results 
under the condition of the local virial relation.
}

\def\non{\nonumber}
\def\be{\begin{equation}}
\def\ee{\end{equation}}
\def\bea{\begin{eqnarray}}
\def\eea{\end{eqnarray}}

\begin{document}

\maketitle

\section{Introduction}
The astronomical
objects in our universe such as galaxies and clusters of galaxies
are believed to be formed through the gravitational interactions among the
mass elements composing these objects. 
These objects originated from tiny fluctuations around the almost homogeneous
background achieve  the virialized state through  gravitational
interactions with the time scale as short as  the system's
 free-fall time. 
However, those
 with the scale larger than the size of a typical galaxy
cannot achieve the thermal-equilibrium state, since the time-scale of
the equilibrium becomes much longer than the age of our universe.
These objects are called collisionless, since the collisional effect
of the gravitational two-body encounters are much less effective than that
of the gravitational mixing through the potential oscillation. 

Collisionless self-gravitating systems (SGS) eventually settle down to a
quasi-equilibrium state through the phase mixing and the violent relaxation
processes under the potential oscillation\cite{Lynden67}. 
This quasi-equilibrium state is a prototype of the astronomical
objects such as galaxies and clusters of galaxies. These objects often show
universal profiles in various aspects.

The quasi-equilibrium states for SGSs have been well examined with N-body simulations
both for the cold collapse simulations 
starting from the small virial ratio \cite{Albada82} and for
the cosmological simulations starting from the tiny fluctuations around the
homogeneous expanding background\cite{Navarro96}. 
Here we pay our attention to the case with the cold collapse and examine the 
common characters of the bound state through N-body simulations.  
Since the cold collapse is considered to be a typical case 
which causes the violent relaxation,
it is meaningful to examine this typical  case in order to grasp the
universal character of violent relaxation  as a first step.

When the system
experiences violent gravitational processes such as cold collapse and
cluster-pair collision, we obtained a universal velocity distribution
profile expressed as the democratic (=equally weighted) superposition of
Gaussian distributions of various temperatures(DT distribution
hereafter) 
in which the local temperature $T(r)$ is defined 
using the local velocity variance $\langle v^{2}\rangle $ as 
$T(r):=m\langle v^{2}\rangle (r)/3k_{B}$, where $m$ is the particle mass
\cite{Osamu04}. Moreover, we have found that the locally defined
temperature linearly falls down in the intermediate cluster region outside
the central part, provided it is described against the cumulative mass $M_{r}
$, \textit{i.e.,} $dT/dM_{r}=const.$ This fact is consistent with the
appearance of DT distributions.

In addition to the linear TM relation, we have also obtained another
peculiar fact that the LV relation between the locally defined
potential energy and kinetic energy holds except for the weak fluctuations,
that is, the local temperature $T(r)$ is proportional to the local
potential $\Phi (r)$, with constant proportionality $6k_{B}T(r)=-m\Phi (r)$
for a wide class of cold collapse simulations  \cite{Osamu04,Sota04}. 

In sec.\ref{sec:isotropic}
we will  show that the LV relation  for SGS 
is  supported by the results of a variety of cold collapse simulations
and  that
among the spherical and isotropic models, 
polytropes with index $n \sim
5$ with a flat boundary condition at the center
are compatible with the numerical results except for the outer region
of the bound state.
In sec.\ref{sec:anisotropic}, we examine the anisotropic models
under the condition of the  LV relation. We will show that the
analytical solutions exist under the LV  condition with constant
$\beta$. These analytical solutions can be utilized to form
the simple model with the anisotropic velocity dispersion. 
Sec.\ref{sec:conc} is devoted to the summaries and
conclusions of this paper.

\section{LV relation and analysis with isotropic models}
\label{sec:isotropic}
In the previous paper, we showed that linear TM relation is one of the
remarkable characters of SGS bound states  after a cold collapse \cite{Osamu04}.
Here we pay our attention to another characteristic for SGS bound states,
that is, the LV relation. 

It is well known that the gravitationally bound system
approaches a virialized state satisfying the condition 
\begin{equation}
\overline{W}+2\overline{K}=0,  \label{Virial}
\end{equation}
where $\overline{W}$ and $\overline{K}$ are, respectively, the averaged
potential energy and kinetic energy of the whole bound system. This is a
global relation that holds for the entire system, after the initial coherent
motion fades out.

Here we define the locally averaged potential energy and kinetic energy
inside the radius $r$, respectively as

\begin{eqnarray}
\overline{W}_{r} &\equiv& \frac{1}{2}\int_{0}^{r}\Phi (r^{\prime })
\rho(r^{\prime })4\pi r^{\prime 2}dr^{\prime },  \label{poteall} \\
\overline{K}_{r} &\equiv& \int_{0}^{r}\frac{\langle v^{2}\rangle (r^{\prime})}{2}
\rho (r^{\prime })4\pi r^{\prime 2}dr^{\prime },  \label{kineall}
\end{eqnarray}
where $\star (r^{\prime })$ means the local object $\star $ evaluated at $%
r^{\prime }$.

Then we extend the virial relation (\ref{Virial}) locally as 
\begin{equation}
\overline{W}_{r}+2\overline{K}_{r}=0,  \label{localvir}
\end{equation}
and examine how precisely this relation is locally attained inside a bound
state.

\begin{figure} 
\begin{center}
\includegraphics[width=8cm]{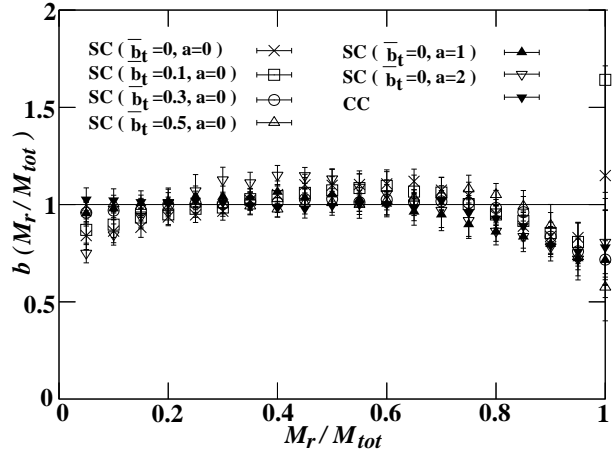}
\end{center}
\caption{ The LV relation for some numerical simulations
obtained by a typical cold collapse simulation;
spherical collapse (SC) and cluster-pair collisions (CC). In the case of SC, 
$5000$ particles are distributed with a power law density profile ($\protect%
\rho\propto r^{-a}$) within a sphere of radius $R$ and the initial virial
ratio ($\bar{b}_t$) is set to be small. In the case of CC, each cluster has
the equal number of particles ($2500$) and all particles are homogeneously
distributed within a sphere of radius $R$ and is set to be virialized
initially. The initial separation of the pair is $6R$ along the $x$ axis. In
all of the simulations, softening length $\protect\epsilon=2^{-8}R$ is
introduced to reduce the numerical error caused by close encounters.
The local virial ratio $b$ is plotted as a function of $%
M_{r}/M_{tot}$, where $M_{tot}$ is the total mass of the system.. 
The virial ratios at $M_r$ are time averaged from $t=5t_{ff}$
and $t=100t_{ff}$. }
\label{LV}
\end{figure}

More precisely, the above relation Eq.(\ref{localvir}) is equivalent to the
purely local relation 
\begin{equation}
2\langle v^{2}\rangle (r)=-\Phi (r),  \label{localvir2}
\end{equation}
at each position $r$. Hence, we define the LV ratio 
\begin{equation}
b(r) \equiv -2 \langle v^{2}\rangle (r)/\Phi(r) ,  \label{localvirr}
\end{equation}
and examine the value of $b(r)$ for each shell. For a wide class of
collapses including cluster-pair collision \cite{Osamu04}, the
value $b(r)$ takes almost unity: it deviates from unity less than 10 percent
upward in the intermediate region and a little more downward in the inner
and outer region for all of the simulations (Fig.\ref{LV}). Hence a wide
class of cold collapse simulations yield the relation (\ref{localvir2})
quite well. This LV relation should be another  characteristic of
SGS, in addition to the linear TM relation.

\begin{figure} 
\begin{center}
\includegraphics[width=8cm]{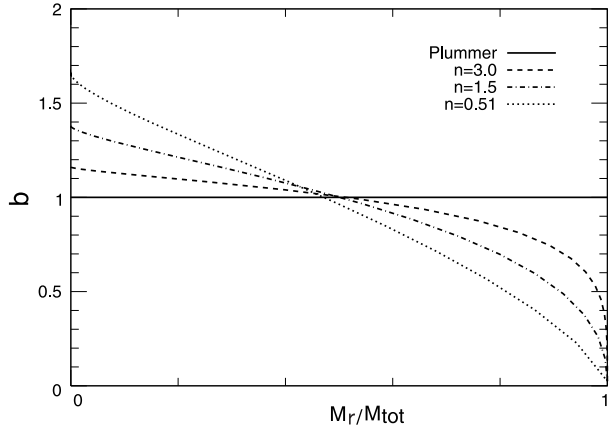}
\end{center}
\caption{ LV relation for polytropes. The solid curve refers to the LV ratio 
$b(M_r/M_{tot})$ as a function of cumulative mass $M_r/M_{tot}$ for Plummer's model.
The other curves refer to $b(M_r/M_{tot})$ for polytropes 
with $n=3.0$ (dashed), $n=1.5$ (dot-dashed), and $n=0.51$ (dotted).}
\label{polyLV}
\end{figure}

From a viewpoint of these two characteristics, we compared our results with 
polytropes. The polytrope with $n=5$  is
special among those models, since it exactly satisfies the LV relation (Fig.\ref{polyLV}).
Moreover, it admits the analytical solution called Plummer's model
under the condition that
the potential is flat at the center of the system \cite{Binney87}. 
We got the result that
Plummer's model  satisfies the linear TM
relation quite well in the intermediate region $0.2M_{tot}<M_{r}<0.8M_{tot}$ \cite{Sota04}. 
On the other hand, both
its TM relation and density profile in the outer region differ from the
ones derived in cold collapse simulations: the temperature falls off more
steeply and the density behaves as $r^{-5}$ in the outer region.

\section{Models with anisotropic velocity  dispersion}
\label{sec:anisotropic}
\begin{figure} 
\begin{center}
\includegraphics[width=8cm]{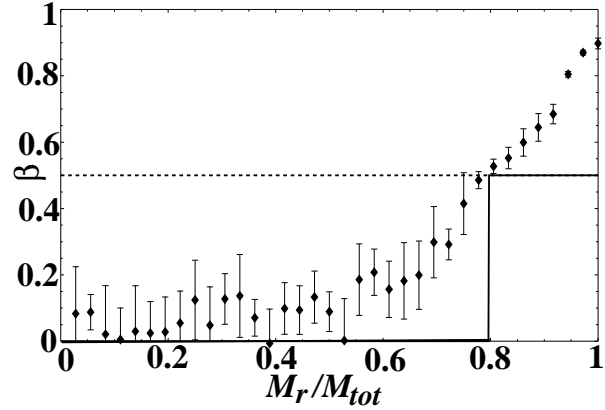}
\end{center}
\caption{ The distribution of the anisotropy parameter $\beta(r)$.
 A plot with error-bar
is the result of the numerical simulation SC with $(\bar{b_t},a)=(0,0)$ 
but with $N=10000$ in Fig.\ref{LV}.
A solid thick line represents the combined solution (\ref{combphi}) 
with $c=0.2$ and a dot line represents the analytical solution
with $\beta=0.5$.}
\label{bmcomb}
\end{figure}

In the previous section, we commented that Plummer's model
 expresses the bound state quite well after a cold collapse
among the spherical systems with isotropic velocity dispersion. However,
it fails in explaining the outer region of the bound state, where the velocity
dispersion is anisotropic (Fig.\ref{bmcomb}). Here we pay our attention to this anisotropy
and examine the general static Jeans equation admitting this anisotropy.

In the collisionless system, the particle distribution can be described 
by the Vlasov equation. 
Integrating the spherical and static Vlasov equation
 in velocity space after multiplying $v_r $ results in the spherical and static
Jeans equation,
\begin{equation}
\label{eq4}
\frac{d\left( {\rho \left\langle {v_r^2 } \right\rangle } \right)}{dr}=\rho 
\frac{d\phi }{dr}-\frac{2\left\langle {v_r^2 } \right\rangle \beta \rho 
}{r},
\end{equation}
where 
$\beta \equiv 1
-\big(\big\langle {v_\theta ^2 } \big\rangle 
+\big\langle {v_\phi ^2 } \big\rangle\big)/
\big(2\big\langle {v_r^2 }\big\rangle\big)$ 
is the anisotropy parameter and $\phi$ is the 
relative potential defined as $\phi \equiv \Phi_{\ast}-\Phi$, where 
$\Phi_{\ast}$ is the maximum energy of the particles in the system \cite{Binney87}.

Assuming  that the density is non-zero everywhere but the
total mass is finite, we can get  the condition that 
$\Phi_{\ast}=0$.
In this case,
from the condition that
$\left\langle {v^2} \right\rangle = \left( {3-2\beta } \right)\left\langle {v_r ^2} 
\right\rangle$,
the LV relation  (\ref{localvir2}) takes the form of 
\begin{equation}
\label{eq5}
\left\langle {v_r ^2} \right\rangle =\frac{\tau }{2}\phi, 
\end{equation}
where $\tau \equiv 1/(3-2\beta)$ .
From Eqs. (\ref{eq4}) and (\ref{eq5}), we derive the relation
\begin{equation}
\label{eq6}
\frac{d\left( {\tau \rho \phi } \right)}{dr}=2\rho \frac{d\phi 
}{dr}-\frac{\left( {3\tau -1} \right)\rho \phi }{r}.
\end{equation}

The relative potential also follows the Poisson's equation 
\begin{equation}
\label{eq7}
\frac{1}{r^2}\frac{d}{dr}\left(r^2\frac{d\phi}{dr}\right)
=-4\pi^2G\rho.
\end{equation}
Eqs.(\ref{eq6}) and (\ref{eq7}) are basic equations for $\phi$ and $\rho$
when the anisotropy parameter $\beta $ is given. 

The equation (\ref{eq6}) is integrable under the condition that $\beta $ is 
constant, which  leads to the relation
\begin{equation}
\label{eq14}
\rho =A\phi ^{5-4\beta} r ^{-2\beta},
\end{equation}
where $A$ is an integral constant.
Substituting Eq.(\ref{eq14}) into Eq.(\ref{eq7}), 
we can get the nonlinear equation for $\phi$, which admits the 
analytical solution 
\begin{equation}
\label{eq24}
\phi \left( r \right)=\frac{G M_{tot}  }{ \left( {r^s+r_0^s} \right)^{1/s}},
\end{equation}
where $s \equiv 2\left( {1-\beta } \right)$ and 
$r_0\equiv\big(\frac{4\pi GA}{s+1}\big)^{1/s}M_{tot}^{2}$
under the boundary condition at infinity;
\bea
& &  \phi \left( \infty \right)=\rho \left( \infty \right)=0 \non\\
& &  {\frac{d\phi }{dr} \Big|_{r \rightarrow \infty} 
     = -\frac{GM_{tot} }{r^2}}
\eea
 \cite{Evans05}.
We will analyze more general  solutions for  eq. (\ref{eq4})
and (\ref{eq7}) under the constant LV ratio $b$
in another paper \cite{Osamu05}. 

\begin{figure} 
\begin{center}
\includegraphics[width=8cm]{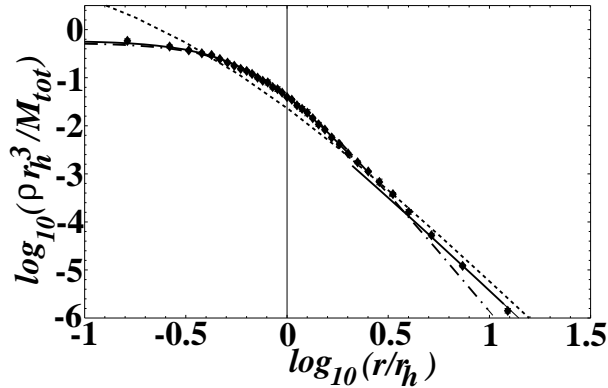}
\end{center}
\caption{ A log-log plot for a  density profile with the unit
of $r_{h}=M_{tot}=G=1$, where $r_{h}$ is the half-mass radius
of the bound system. A plot with error-bar
is the result of the numerical simulation SC with $(\bar{b_t},a)=(0,0)$ 
but with $N=10000$ in Fig.\ref{LV}.
A solid line represents the combined solution (\ref{combphi}) with $c=0.2$.
A dot  and  dot-dashed line represent  the analytical solution
with $\beta=0.5$ and the Plummer's model, respectively.
The coincidence between numerical data and the combined solution is
remarkable. }
\label{rhocomb}
\end{figure}

\begin{figure} 
\begin{center}
\includegraphics[width=8cm]{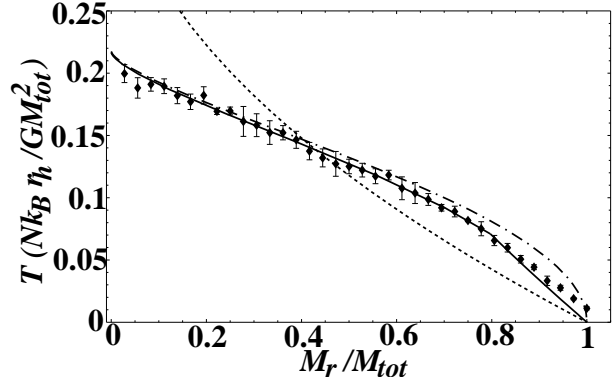}
\end{center}
\caption{ A Temperature-Mass relation 
with the unit of $r_{h}=M_{tot}=G=1$, where $r_{h}$ is the half-mass radius
of the bound system. A plot with error-bar
is the result of the numerical simulation SC with $(\bar{b_t},a)=(0,0)$ 
but with $N=10000$ in Fig.\ref{LV}.
A solid line represents the combined solution (\ref{combphi})  with $c=0.2$.
A dot  and  dot-dashed line represent  the analytical solution
with $\beta=0.5$ and the Plummer's model, respectively.
The coincidence between numerical data  and the combined solution is
remarkable. }
\label{tmcomb}
\end{figure}

Especially when $\beta=0$, the solution (\ref{eq24}) is nothing but
Plummer's model. 
The Plummer's model explains several characters of the bound state
after a cold collapse or cluster-pair collisions quite well 
except for the outer region. 
Here we compare this solution with the solution connecting
inner Plummer's solution with the outer analytical solution with
constant $\beta$ (See Appendix \ref{appe1}). Of course, the actual function form of $\beta(r)$ is
too complicated to be expressed by the connection of several constant
$\beta$ analytical solutions. However, we can roughly see that considering
the anisotropy reduce the incompatibility of the solution to the bound state
after a cold collapse (Fig.\ref{rhocomb} and Fig.\ref{tmcomb}).

\section{Conclusions} 
\label{sec:conc}
The bound states after a cold collapse or 
several other initial conditions 
with small initial virial ratio are characterized 
by the linear temperature-mass (TM) relation
and the local virial (LV) conditions. 
Among the spherical and static models as the solution of Vlasov equation
with isotropic velocity dispersion, the polytrope with $n=5$ is very special,
since it satisfies the LV relation exactly. Plummer's solution, which
is the analytical solution of polytrope with $n=5$ satisfying the flat boundary
condition at the center explains several characters of the bound state
quite well except for the outer region. Here we showed that this incompatibility
can be reduced by adding the anisotropic effect under the condition of 
the LV relation. This leads to the result that
 the bound state after a cold collapse characterized
by the density profile with the asymptotic form $\propto r^{-4}$ and
linear TM relation can be explained quite well by the anisotropic model
satisfying the LV relations. 

Here we have restricted our analysis 
to  cold collapse  models with small initial virial
conditions. However,
 it  seems important how widely our results  are applicable in more
general mixing processes including  merging process in cosmological simulations.
It is well known that the bound states observed in cosmological simulations
are quite different from those in cold collapse.
For example, the density has the central cusp  $\propto r^{-1}$ and takes
the outer asymptotic form $\propto r^{-3}$ in cosmological simulations \cite{Navarro96}.
Moreover, the phase space density defined as $\rho/\sigma^3$
has a single power-law property 
in radius  \cite{Taylor01}. 
Hence it seems important to see if the LV relation is admitted even in
the case with such different characters. 
These points are
now under investigation \cite{Osamu05}.


\appendix
\section{}
\label{appe1}

Here we connect the two analytical solution  at $r=r_c$., i.e., Plummer's model ($\beta=0$) for  $r \leq r_c $
and
$\beta=0.5$ analytical solution  for  $r \geq r_c $.
Defining $\phi _c \equiv \phi \left( {r_c } \right)$ and describing the 
physical quantities with the unit of 
$r_c =\phi _c =G=1$,
the connected solution is described with one-parameter $c$ as 
\be
\phi \left( r \right)=\left\{ {{\begin{array}{*{20}c}
 {\sqrt {\frac{1+3c^2}{r^2+3c^2}} } \hfill & {\left( {r\le 1} \right)} 
\hfill \\
 {\frac{1+3c^2}{r+3c^2}} \hfill & {\left( {r>1} \right)} \hfill \\
 \end{array} }}\right.
\label{combphi},
\ee
from the continuity condition of 
$\phi(r)$ and ${d\phi(r)/dr}$ at r=1. 

Inner Mass $M(r)=r^2 d\phi/dr$ becomes
\be
M\left( r \right)=\left\{ {{\begin{array}{*{20}c}
 {\frac{\sqrt {1+3c^2} r^3}{\sqrt {\left( {r^2+3c^2} \right)^3} }} \hfill & 
{\left( {r\le 1} \right)} \hfill \\
 {\frac{\left( {1+3c^2} \right)r^2}{\left( {r+3c^2} \right)^2}} \hfill & 
{\left( {r>1} \right)} \hfill \\
\end{array} }} \right..
\label{combmass}
\ee
Hence the total mass $M_{tot} $ is described as 
\[
M_{tot} =\mathop {\lim }\limits_{r\to \infty } M\left( r \right)=1+3c^2.
\]
The mass ratio between the total mass of inner Plummer region, $M_{plum} $ 
and $M_{tot} $

becomes
\[
\frac{M_{plum} }{M_{tot} }=\frac{1}{\left( {1+3c^2} \right)^2}.
\]
Hence the parameter $c$ determines the mass ratio of the connected two 
regions. 

%

\end{document}